\documentclass[a4paper,notitlepage,12pt]{article}
\usepackage{amsfonts}
\usepackage{amssymb}
\usepackage{amsmath}
\usepackage{bm}
\usepackage{graphicx}
\usepackage{multirow}
\usepackage{color}
\usepackage{graphics}
\usepackage{cite}

\def\one{{\hbox{1\kern-.8mm l}}}

\newcommand{\beq}{\begin{equation}}
\newcommand{\eeq}{\end{equation}}
\def\be#1\ee{\begin{align}#1\end{align}}
\newcommand{\ov } {\over }

\begin{document}
\begin{titlepage}
\bigskip\begin{flushright}
NORDITA-2010-18\\
\end{flushright}
\begin{center}
\vskip 2cm
\Large{Gravitational waves from first order phase
  transitions during inflation}
\end{center}
\vskip 1cm
\begin{center}
\large{Diego Chialva \\ 
       {\it  Nordita Institute,
       AlbaNova University Centre,
   Roslagstullsbacken 23
   SE-106 91 Stockholm, Sweden} \\
        \tt{chialva@nordita.org}}

\end{center}
\date{}

\pagestyle{plain}

\begin{abstract}
We study the production, spectrum and
detectability of gravitational waves in models of the early Universe
where first order phase transitions occur during inflation.

We consider all relevant sources. The self-consistency of
the scenario strongly affects the features of the waves.
The spectrum appears to be mainly sourced by collisions of
bubble of the new phases, while
plasma dynamics (turbulence) and the primordial gauge fields connected to the
physics of the 
transitions are generally subdominant. 

The amplitude and frequency dependence of the spectrum for modes that
exit the horizon 
during inflation are different from those of the waves produced by 
quantum vacuum oscillations of the metric or by
first order phase transitions not occurring during inflation.

A not too large number of slow (but still successful) phase transitions can
leave detectable marks in the CMBR, but the signal weakens rapidly for
faster transitions. When the number of phase transitions is instead large, the
primordial gravitational waves can be observed both in the CMBR or
with LISA (but in this case only marginally, for the slowest
transitions) and especially with DECIGO.

We also discuss the nucleosynthesis bound and the constraints it
places on the parameters of the models. 
\end{abstract}


\end{titlepage}

\tableofcontents

\setcounter{section}{0}

\section{Introduction}

Gravitational waves
carry valuable information about the physics that produced them, as
they decouple quite soon from their surrounding. In 
particular, waves generated during inflation could 
open open up important opportunities to study the
early Universe.

Beside the ever present generation in vacuum via quantum
fluctuations, gravitational waves can also be sourced by the
anisotropic  
stress tensor of fields and fluids.
More specifically, important sources are expected to be present when
first order phase transitions occur.
In the literature, there has been great interest in the
gravitational waves generated by this kind of transitions (see
\cite{preheat, ew, Kosowsky:1992vn, 
Kamionkowski:1993fg, Huber:2008hg, Caprini:2009fx, 
Caprini:2007xq}). The analysis, however, has been 
mostly concerned with transitions such as
the electroweak one\footnote{Which however does not appear to be a first
  order transition for the present lower bound on
  the Higgs mass.} \cite{ew} or during preheating \cite{preheat}. 

In this work we instead investigate the production, features and detectability
of gravitational waves from
 first order phase transitions during inflation.

Inflationary models exhibiting this kind of transitions have
existed since  
the early times of inflationary theory: for example Guth's Old
Inflation was indeed driven by a first order phase
transition. However, the motivation for this analysis is even stronger today,
because of the appearance of many metastable 
vacua in effective theories of gravity and high energy physics,
where tunnelings and 
transitions among vacua are expected to occur, possibly during inflation.
For example, it has been shown that long series of connected minima can
exist in the string theory landscape \cite{Danielsson:2006xw}. 

Furthermore, recent investigations have fully analyzed
\cite{Chialva:2008zw} a cosmological model alternative 
to slow-roll/chaotic inflation, where the inflationary dynamics is actually
driven by several first order phase transitions: chain inflation.  

Although expected to be important, the signatures in gravitational
waves from these scenarios and 
models have not been studied: this work intends to fill this
gap\footnote{A partial 
   analysis of waves emission in the specific setup of Fourth Order
   Gravity (FOG) was 
   done in \cite{Baccigalupi:1997re}. Those 
  results and approach are different from ours.
In \cite{Ashoorioon:2008nh} an analysis of gravitational waves in a specific
realization of chain inflation was
attempted. Our results are very different from those, as in
\cite{Ashoorioon:2008nh} the analysis appears not to be consistent.
The work is not published on a scientific journal.}. 

This paper is organized as follows: in section \ref{overview} we
present the setup and the approach we follow. The
detailed investigation is then divided in sections
\ref{parameterssec}, \ref{emission}, \ref{discussionconclusion}.
We start in section \ref{parameterssec} by studying our setup in
terms of the relevant physical parameters and
the bounds on them coming from the request of self-consistency of the scenario
and its description. These bounds will affect also the emission and
characteristics of the gravitational waves.

In section \ref{emission} we analyze the production of the
gravitational waves and their features. We study various possible
sources related to first order phase transitions during inflation. 
More specifically,
in section \ref{collisionswaves}
we compute the spectrum of waves
emitted by the collisions of bubbles, which
appear to be the strongest source.
In section \ref{fluidwaves}, we study the remaining ones.

Finally, in section \ref{discussionconclusion} we
discuss the detectability of the waves in the CMBR and at interferometers.

We conclude in section \ref{discussion}. The appendices contain
useful accessory material.

\section{Setup and approach}\label{overview}

Consider a period of inflation in the early Universe where
some first order
phase transitions occur. The theory describing this scenario could be
very complicated, with a potential exhibiting many metastable minima
at different energies, a large number of fields and a complex dynamics, 
with rolling, tunneling and jumping phases as the
fields pass through the minima.

The inflationary dynamics could occur in various ways depending on the
behavior of the fields: for example via
the mechanism of chain inflation \cite{Chialva:2008zw}, or when one
scalar field undergoes slow-rolling while others tunnel through
the minima.

The first order phase transitions take place via nucleation of bubbles
of the new phases 
within the old ones. With the expansion of the bubbles and their
collisions, the latent heat of the transitions\footnote{In 
  the following, with the
  term transition we will
  always intend a first order one, even if we do not explicitly write
  it to avoid 
  repetitions.} is released and converted in a
radiation-dominated fluid. Many sources of gravitational waves become
active due to these dynamics.

We will simplify the description of this setup.
In fact, knowledge of the details of the field theory is not
necessary for the kind of 
analysis we are going to perform. Inflation and the phase
transitions can be described by a series of physical
parameters (for example, the time-scales of the transitions, the
Hubble parameter, 
the nucleation size of bubbles and a few others
to be introduced in 
section  \ref{parameterssec}). We will study the production and features of the 
gravitational waves using these
parameters, without resorting to the complicated field theory description.

The analysis is nevertheless quite complex:
instead of computing the physical parameters from first principle via
the field theory, we will constrain them on the basis of the
requirements of consistency of the scenario. Indeed, the phase
transitions must not backreact too strongly on the background, if we
want inflation not to be stopped and to be efficient. At the same time
they must be successful, reaching percolation and large scale thermalization,
in spite of the fast expansion of the Universe.

All of this also affects the physical parameters of the
sources of gravitational waves, constraining the
amplitude and features of the latter. The constraints are so
binding, that we will be
able, using our analysis, to fully determine if a series of first
order phase transitions compatible with inflation can produce a
detectable spectrum of gravitational waves. 

An advantage of the approach we use is that the results are of a
general nature and can be adapted to
many different field theory models. Indeed, once a particular model is chosen,
one can compute from first principles the physical parameters,
which will now depend on the couplings of the fields and the dynamics of
the model, and specialize our results on gravitational wave emission
to the case of interest, being able to test it. We will do this, in
the end, using the example of chain inflation.

\section{Analysis of the setup: parameters and bounds}
\label{parameterssec}

We need to consider a limited number of physical parameters to analyze
the emission and features of the gravitational waves. We also need to
investigate which of their values allow successful transitions
together with efficient
inflation, homogeneity and isotropy. 

In the following we will use units for which $c=\hbar=1$.
We also use $t$ for the cosmic time, $\eta$ for the 
conformal and define
 $
  M_{\text{Planck}}^2=(8\pi \, G)^{-1}
 $
.
The\, $\dot{}\, (')$\, indicates derivative by $t\,(\eta)$.

\subsection{Evolution of the background}
We need only two parameters:
\begin{itemize}
 \item The Hubble scale $H = {\dot a \ov a}$, here written in cosmic
   time ($a$ is the scale factor). 
  \item The parameter $\varepsilon= - {\dot H \ov H}$ indicating the
    time evolution of the Hubble scale. We consider quasi-de Sitter
    scenarios, where
    $\varepsilon < 1$ during inflation.
\end{itemize}

These parameters come from using a Friedman-Robertson-Walker metric
for the background (see appendix \ref{basic}). The latter 
is appropriate if there are homogeneity and isotropy at least
above certain scales, that is 
if the bubble of the phase transitions do reach
percolation and 
large scale thermalization, and if the radiation-fluid generated by
the collisions also thermalizes. We
are soon going to discuss these points.

\subsection{Phase transitions}
We take into account the possibility that more than one phase
transition occur, by indicating each of them
with a progressive integer $1 \leq n \leq N$; $N$ is kept
generic. 

We list here the relevant physical parameters, postponing the
discussion of their bounds to section \ref{conditionsdecayrates}:
\begin{itemize}
 \item the decay rate
    $\widetilde\Gamma_n$  
   per unit time between phases $n$ and $n-1$ (also called
   nucleation, or tunneling rate). 

  $\widetilde\Gamma_n$ is related to the decay rate per unit time and
   volume $\Gamma_n$, which is the quantity usually obtained in a
   field theory model via tunneling action (or free energy if
   the temperature is important) \cite{Linde:2005ht}: 
   $\widetilde\Gamma_n = \int dV_{\text{phys}} \Gamma_n$, where
   $V_{\text{phys}}$ is the physical volume.

 \item The time-scale $\beta_n^{-1}$ of the phase
   transition $n \to n-1$.

   $\beta_n^{-1}$ is the
   lapse of cosmic time during
   which most of the bubble nucleate, collide and thermalize.
   In 
   appendix \ref{timescaletrans}, we show its relation with the
   decay rate and the tunneling action. The time-scale in conformal
   time is indicated with $\tilde\beta_n^{-1}$ and defined in
   \ref{timescaletrans}. 

 \item the energy density $\Delta \epsilon_n$  released by the
   transition among phases $n, n-1$. 

  In each transition some energy is liberated.
  The energy density at disposal is
  $
   \Delta \epsilon_n \equiv \epsilon_n-\epsilon_{n-1} \, ,
  $
  where $\epsilon_m$ indicates the energy density in the phase $m$. 
  It is carried by the bubble walls and transferred by 
  the transition and the collisions to the fluid velocity and heating,
  and ultimately to the
  perturbations such as the gravitational waves.

 \item The nucleation size $r_{n}$ of the bubbles of the
   transition between phases $n, \, n-1$.
 
   We want to express $r_{n}$ in terms of other physical parameters.
   To proceed, we need to know a bit more about the
   process of bubble nucleation. 
 
   The growth or not of a bubble
   can be seen as a competition between the expansion due to the
   release of energy from the transition and the surface tension of
   the bubble wall\footnote{We neglect
     gravity as it will be easy to check later that $r_n \ll
     R_S$, where $R_S \sim (G \Delta\epsilon_n)^{-{1 \ov 2}}$ is the
     gravitational radius signaling the need to consider gravity
     ($G$ is newton's constant)
     \cite{Coleman:1980aw}.}. Only bubbles 
   of an initial size larger or equal than a 
   certain critical value can effectively be 
   nucleated and grow. 

   A precise
   description is possible in terms of a tunneling action, or
   free energy in case the temperature is not zero, for the order
   parameter of the transition \cite{Linde:2005ht}. The tunneling
   action/free energy is
   indeed the sum of a term relative to the bubble's wall tension
   minus a term for the energy of the bubble interior. The critical
   radius is computed by extremizing it.

   In the thin wall approximation for the tunneling, we find
   \be \label{nucleationsize}
    r_{n} & =  {3 S_n \ov \Delta\epsilon_n}=
    \left({2 \ov 27\,\pi^2}\right)^{{1 \ov 4}}
    \left({S_E^{(n)} \ov \Delta\epsilon_n}\right)^{{1 \ov 4}} &
        & \text{for $T < r_{n}^{-1}$} \\
    r_{n} & =  {2 S_n(T) \ov \Delta\epsilon_n(T)}=
    {2 \ov (16\,\pi)^{{1 \ov 3}}}
    \left(F_E^{(n)}{T \ov \Delta\epsilon_n}\right)^{{1 \ov 3}}&
        & \text{for $T > r_{n}^{-1}$} \, ,
   \ee
   where $S_n$ is the surface tension of the bubble wall and
   $\Delta\epsilon_n$ has been defined above.
   We have also expressed the critical radius in terms of the extremized
   tunneling action $S_E^{(n)}$ or free energy
   $F_E^{(n)}$, where
   $
   S^{(n)}_E  = {27 \pi^2 S_n^4 \ov \Delta \epsilon^3} \quad
   F^{(n)}_E  = {16 \pi S_n(T)^3 \ov T \, \Delta \epsilon^2 }
   $\, \cite{Linde:2005ht}.

   The two formulas above are respectively valid for vacuum or thermal
   nucleation\footnote{Recall that the latter could be possible, as after the
     first transition completes there is also a radiation component of
     the Universe coming from the bubble walls decay.}. The tunneling
   action is used if \, $T < r^{-1}_{n}$ (vacuum description),
   otherwise the free energy must be employed \cite{Linde:2005ht}.
   We will show in section \ref{conditionsonT} which of the two
   description is more appropriate. 
\end{itemize}

\subsubsection{Conditions on the decay rates and the time-scales of
  the transitions}\label{conditionsdecayrates}

The phase transitions must be successful, reaching the
stages of percolation, 
bubble collisions and large scale thermalization, preserving
the homogeneity and isotropy of the Universe. In order to do so, the
transition must cope with the inflationary expansion of the Universe.

This puts a series of conditions on the decay rate $\widetilde
\Gamma_n$ and the time scale $\beta_n^{-1}$ of the transition
\cite{Chialva:2008zw, Turner:1992tz}. In particular, the necessary
conditions for 
successful completion of a transitions during inflation is
\cite{Chialva:2008zw} 
 \beq \label{successfulcompletion}
  \widetilde \Gamma_n > 3H \, .
 \eeq
$\widetilde\Gamma_n$ depends in general on time\footnote{For example  
through the dependence of the
tunneling action on different fields.}: we can
have {\em fast tunneling models} (as chain inflation), where the decay 
rate is large enough at the onset for the transition to 
complete very rapidly, and 
{\em evolving tunneling} models, where
the transition occur
initially with a smaller rate (the Universe is trapped in the old
vacuum for a certain time) and finishes when the tunneling rate
has increased enough to satisfy (\ref{successfulcompletion}). 

However, to preserve homogeneity and isotropy, most of the
bubbles must nucleate and collide in a short time, compared to the
Hubble time (big bubbles are dangerous, see
\cite{Turner:1992tz}). This implies
 \beq \label{lowerboundtimescale}
  \beta_n^{-1} < H^{-1} \, .
 \eeq

We are now going to find more stringent bounds on $\beta_n$
considering the evolution of the bubbles.

Bubbles are nucleated with radius $r \sim r_n$.
By the time of collision, a point on the surface
of the wall of a bubble has moved by a 
physical distance 
 \beq \label{finalsize}
  r_{f, n}-r_{n} = {v \ov H}\left(e^{{H \ov \beta_n}}-1\right)
  \, ,
 \eeq
where $v$ is the wall velocity. We have assumed $H$ and $v$ to be
constant during 
the time of the transition, as the time-scale of the latter is
necessarily short, as we said. 
Initially the bubbles are spherical.  We assume, again because of
the short evolution time-scale, that they retain that shape.

In all realistic cases of successful transitions
two more conditions are verified\footnote{In principle, we could
  partially relax these conditions, but those transitions would not be
  typical and, more 
  importantly, the existing numerical studies \cite{Kosowsky:1992vn, 
  Kamionkowski:1993fg, Huber:2008hg, Caprini:2009fx}, which we
  extend in
  section \ref{collisionswaves}, would not be applicable.}
\cite{Kosowsky:1992vn,  
  Kamionkowski:1993fg, Caprini:2007xq, Huber:2008hg, Caprini:2009fx,
  Turner:1992tz} 
 \be
  r_{f, n}-r_{n} & \sim  {v \ov \beta_n} \label{flatbubblesizeevolution}\\
  r_{n} < {v \ov \beta_n} \label{smallinitialsize}\, .
 \ee
For (\ref{finalsize}) and (\ref{flatbubblesizeevolution}) to
be consistent at least 
to an acceptable value (say 5\%), it must be
 \beq \label{lowerbetastrong}
  {\beta_n \ov H} \gtrsim 10 \, .
 \eeq
Condition (\ref{smallinitialsize}), instead, gives an
upper bound on $\beta_n$. We start by writing
$\Delta \epsilon_n$ as
 \beq \label{energybound} 
  \Delta \epsilon_n \sim - {d \rho \ov dt} \beta_n^{-1} \simeq 6 H^3
  M_{\text{Panck}}^2 \beta_n^{-1} \varepsilon \, ,
 \eeq
where $\rho$ is the total energy density dominated by the
vacuum component, and we have used the Friedman
equation.

Inserting (\ref{energybound}) in
(\ref{nucleationsize}), we find from (\ref{smallinitialsize})
 \beq 
  {\beta_n \ov H} < 
   \left({\pi^2 \ov S_E^{(n)}}\right)^{{1 \ov 5}}10^{{8 \ov 5}}\, .
 \eeq
We have considered only the case $T < r_n^{-1}$ as we will soon show
that it is the relevant one.

Summarizing, the bounds on the ratio between the scale of the
transitions and the Hubble rate are
 \beq \label{sintbound}
  10 \,\, < \,\, {\beta_n \ov H} \,\, < \,\,
   \left({\pi^2 \ov S_E^{(n)}}\right)^{{1 \ov 5}}10^{{8 \ov 5}}
   \, .
 \eeq

\subsection{Radiation fluid}

The collision of walls and the release of the latent heat generically
produce a radiation-dominated fluid\footnote{We will show that it is
  relativistic (radiation) in section \ref{conditionsonT}.}. In order
to have an effective 
Friedman-Robertson-Walker description of the metric, the fluid must
thermalize and
the time-scales for the decay of the collided walls
and the fluid thermalization must be short\footnote{If it were
  necessary to remind it: it is well known 
  that in all FRW cosmological models
  matter and radiation are really never in thermal equilibrium, as
  those space-times do not posses a time-like Killing
  vector. However, in general the Universe can be, and often is, very
  near thermal equilibrium and this is the meaning
  of thermalization in this context \cite{KolbTurner}. See also
  \cite{Chialva:2008zw}.}. 

The process of 
bubble collision and the transfer of the energy to radiation are
complex phenomena
that can be studied numerically. We are not going to do so and we will
assume that the thermalization of the fluid occurs rapidly.

Note that the production of radiation cannot be neglected: although during
inflation it is a sub-dominant fraction of the total
energy density in comparison with the vacuum, it can be
important for what concerns the perturbations. This was shown, for
example, in the case of 
chain inflation with a single scalar field, where it was
actually essential to provide a spectrum of adiabatic density (scalar)
perturbations in accordance with observations \cite{Chialva:2008zw}.

The parameters necessary to describe the fluid are the following ones:
\begin{itemize}
 \item The temperature $T$ of the radiation fluid. 

   Patches of the Universe in different phases have in general different
   temperatures, and we should distinguish them with a label $n$. As 
   we will see (section \ref{transitionscollisionsfeatures}), the
   differences between the temperatures of the phases are (and must be)
   negligible for inflation to be efficient, therefore we will often omit
   the suffix $n$. 
 \item Plasma scales.

  The constituents of the fluid will
  in general be 
  charge carriers. Such fluid
  goes under the name of plasma. More refined definition of
  plasma are possible, but we will not consider them.

  When it is relativistic, its dynamics
  is regulated only by the gauge coupling(s) $g$ and the temperature
  $T$. The relevant 
  scales, such as the plasma frequency $\omega_p \sim g T$ or the 
  screening distance $\lambda_D \sim (g T)^{-1}$,   
  are determined by these quantities\footnote{One
    generally distinguishes between 
    different kind of collective oscillations \cite{Arnold:2000dr},
    but their characteristic frequencies are of the same order.
    Also, non-abelian and abelian gauge theories   
  are generally distinguished by different constant of
  proportionality involved in the formulas for the typical scales. We
  treat both cases 
  by indicating the plasma scales up to the proportionality
  constants.} \cite{Arnold:2000dr}.
  
  Other important plasma scales and parameters will be introduced in
  section \ref{fluidwaves}, when studying the gravitational
  emission.
\end{itemize}
 
\subsubsection{Conditions on the fluid temperature}\label{conditionsonT}

A quasi- de Sitter inflationary evolution imposes an upper
bound on the fluid energy density $\rho_f$ and therefore its temperature $T$. 

By using the Chaudhuri equation (\ref{Chaudhuri}) and
the parametrization of the Hubble constant 
 \beq \label{WMAPnormH} 
  H = \sqrt{8 \pi^2 \, c_s \,\eta \, \varepsilon}\,M_{Planck} 
   \sim 10^{-4} \sqrt{\varepsilon}\,M_{Planck} \, ,
 \eeq  
obtained normalizing the spectrum of scalar
  perturbations 
$\sim {H^2 \ov 8 \pi^2 \, c_s \,\varepsilon\,M_{Planck}^2}$ to 
the result of the 5-years  WMAP survey $\Delta _{\mathcal{R}}^{2}=\eta
\sim 2.5\times 10^{-9}$ \cite{WMAP5},
we find
 \beq \label{bathtemperature}
  T = \left({\rho_f \ov \kappa}\right)^{{1 \ov 4}} \,\,
    < \,\, 10^{-2}\kappa^{-{1 \ov 4}}\varepsilon^{{1 \ov 2}}\,M_{Planck} \,,
 \eeq 
where $\kappa=\kappa(T)$ counts the number of relativistic degrees of freedom at
temperature $T$.
The inequality follows from the fact that in the specific models there can
be other components entering the Chaudhuri equation beside
radiation, for  example the kinetic energy of scalar field(s).

If $\varepsilon$ is not
pathologically small, $T \gg GeV$.  
At such temperature
particles\footnote{We do not know the theory at those energies, it
  could be supersymmetric or also a GUT.} are effectively
massless. 

Although the temperature is constrained as shown in
(\ref{bathtemperature}), we still have to check the modification to
the tunneling dynamics due to thermal corrections. In fact, 
the transitions (tunneling) and the gravitational emission can change
depending on the temperature.
The vacuum tunneling description is well-suited
to capture the dynamics only  
provided that the temperature  $T$ is smaller than the inverse of the
scale length of the field solution $r_{n}^{-1}$ \cite{Linde:2005ht}.
From (\ref{nucleationsize}), (\ref{sintbound}),
(\ref{bathtemperature}), we find
(for $\kappa \sim 100$) 
 \beq \label{ratioTinverrn}
  {T \ov r_{n}^{-1}} < \left({6 S_E^{(n)}\ov \pi^2}\right)^{{1 \ov 4}}
    \left({\beta_{n} \ov 4 \kappa H}\right)^{{1 \ov 4}} < 1 \quad
  \text{for}\, S_E^{c, (n)} < 20 \, .
 \eeq
Noting that the critical tunneling action cannot be too large as it
will suppress the decay rate too much, also in view of
(\ref{successfulcompletion}), we conclude that 
the pure vacuum description of the
tunneling is appropriate. 

\subsection{Final comments}

The bounds on the typical scale of the transitions
and the temperature (especially
(\ref{sintbound}), (\ref{bathtemperature})) will strongly affect the
gravitational emission and the sources. 

Let us anticipate that, in particular, the physics of the plasma will
prove to be a subdominant source because of the constraint
(\ref{bathtemperature}) on the temperature. The bounds
(\ref{sintbound}) will instead be very important for discussing the
suppression of the 
waves emitted by 
the collisions of bubbles.

Note also that the energy density $\Delta \epsilon_n$ released by a
transition is much smaller than the total energy density, by
a factor $2{H \ov \beta_n} \varepsilon$, see (\ref{energybound}). This
will have important consequences for the gravitational emission.

\section{Gravitational waves}\label{emission} 

We organize the study of the gravitational waves in the following sections. 

\subsection{Useful definitions and notation}\label{generalsolution} 

A gravitational wave can be seen as a ripple on the top of the background
metric, as (in conformal time $\eta$)
 \beq
  ds^2 = a(\eta)^2 [-d\eta^2 + (\delta_{ij}+h_{ij})dx^i\,dx^j]\, .
 \eeq 
In particular, gravity waves are true tensor modes and are transverse, symmetric
and traceless.

The linearized Einstein equation reads\footnote{In cosmological
  perturbation theory 
  the sources we will consider arise at second order. At that level
  tensor perturbations are 
  gauge dependent.}
 \beq \label{gravitywaveseqmot}
  h^{''}_{ij}+2\mathcal{H}h^{'}_{ij}-\nabla^2h_{ij}= 16 \pi G a^2 \pi^T_{ij} \,.
 \eeq
where $\mathcal{H} = a H$ and $\pi^T_{ij}$ is the traceless symmetric
transverse tensor part of the 
anisotropic stress tensor, obtained 
by suitable projection from the total
energy-momentum tensor.

Not indicating polarization indexes 
to avoid cluttering of formulas,
the spectrum $P_h(k, t)$ of gravitational waves is defined as 
 \beq \label{spectrumdef}
  \langle h^*(\vec k, t) h(\vec k', t) \rangle = \delta^{(3)}(\vec k-\vec k')
   {2 \pi^2 \ov k^3} P_h(k, t)
 \eeq
and the energy density per octave
 \beq \label{energyraddef}
  k{d \rho_h(\vec k, t) \ov dk} = k^3\int d\Omega 
   {\langle \dot h^{*}(\vec k, t) \dot h(\vec k', t) \rangle \ov (2\pi)^38\pi G}
    =
    k^3 \int d\Omega
 {\langle h^{*'}(\vec k, \eta) h'(\vec k', \eta) \rangle \ov (2\pi)^38\pi G a^2}
    \, .
 \eeq
Here, $h(\vec k, \eta)$ is the mode function of the graviton
field $h(\vec x, \eta)$ expanded in
eigenfunctions of the Laplace-Beltrami differential operator with eigenvalues
$-k^2$. The wavenumber $k$ is related to the physical momentum by $k =
a(t) p$.

The brackets in (\ref{spectrumdef}), (\ref{energyraddef})
refer to a quantum treatment of perturbations.
We will evaluate the two-point functions on the Bunch-Davies
vacuum, neglecting the so-called transplanckian issue.

\subsection{Formal solution of the wave equation}

In order to discuss the observables we will be interested in, the
field equation (\ref{gravitywaveseqmot}) must be solved both during
inflation and later on, during matter or radiation domination.

For a quasi- de Sitter background during inflation
($a= - {1 \ov \eta H(1-\varepsilon)}$), we find
the general solution 
for the mode functions
 \be \label{hgeneralinhomogeneoussolution}
  h & =  
  \left(c_1 +i{\pi \ov 4}\int^\eta d\eta' (\eta')^{1-\nu^T} H_{\nu^T}^{(2)}(-k\eta') S(\vec k, \eta') \right) \,
  (-\eta)^{\nu^T} \, H_{\nu^T}^{(1)}(-k\eta)  \nonumber \\
   & \quad + 
   \left(c_2-i{\pi \ov 4}\int^\eta d\eta' (\eta')^{1-\nu^T} H_{\nu^T}^{(1)}(-k\eta') S(\vec k, \eta') \right)\,
  (-\eta)^{\nu^T} \, H_{\nu^T}^{(2)}(-k\eta) \, ,
 \ee
where $H_{\nu^T}^{(1, 2)}$ are Hankel functions and
 \beq
  \nu^T = {3 \ov 2}+\varepsilon \sim {3 \ov 2}\, , \qquad
  S(\vec k, \eta)  = 16 \pi G a^2 \,\pi^T (\vec k, \eta)\, .
 \eeq

The terms in the solution proportional to $c_1, c_2$ are related to
the vacuum fluctuations of the field (homogeneous equation). We choose
the Bunch-Davies vacuum setting $c_1 = H{\sqrt{\pi} \ov 2}
e^{i(\nu^T+{1 \ov 2}){\pi \ov 2}}, \quad c_2= 0$.

During matter or radiation domination, when the source is no longer
active and the scale has reentered the horizon ($k \gg \mathcal{H}$), the
solution is
 \beq \label{highfreq}
  h = {A_+(k) \ov a\sqrt{k}}e^{-ik \eta}+{A_-(k) \ov a\sqrt{k}}e^{ik \eta} \, .
 \eeq
The  constants $A_{\pm}(k)$ are determined by properly matching
(\ref{hgeneralinhomogeneoussolution}), (\ref{highfreq}) so that both $h$
and $h'$ are continuous.
The time of matching depends on the duration of inflation and on the
value of $k\eta$, which signals when and if the scale $k^{-1}$ is
inside or outside the horizon.

\subsection{Sources of gravitational waves in the presence of first order
  transitions during inflation}
 
First order phase transitions provide many possible sources for
gravitational waves. In particular we will consider
\begin{itemize}
    \item the anisotropy stress tensor from
      bubble collision 
    \item the velocity spectrum of the fluid
    \item hydrodynamical turbulence
    \item non-zero gauge fields and
      (hyper)magneto- hydrodynamical\footnote{We will indicate with the
      term ``hyper'' the fields associated with a $U(1)$ gauge
      symmetry, although in general that will not be necessarily
      related to the hypercharge of the Standard Model.} 
      turbulence
    \item anisotropy tensor of the radiation fluid (viscosity)
\end{itemize}
In the
  formal expansion of cosmological perturbation theory, these sources
  arise at second order.
At that order, tensor modes are also sourced by first order metric
perturbations,  
coming from the expansion both of the energy momentum
and of the Einstein tensors \cite{Inducedsecondorder}. We will not
discuss this in the present work.

We turn now to the detailed analysis of the sources listed above and their wave
emission during inflation.  

\subsection{From phase transitions and bubble
  collisions}\label{collisionswaves} 

The collisions of bubbles source gravitational waves. 
Instead, the evolution of bubbles prior to collision does not generate
gravitational waves because of the spherical symmetry of the bubbles
\cite{Caprini:2007xq}. 

\subsubsection{Features of the source}\label{transitionscollisionsfeatures}

We have found at the end of section \ref{conditionsonT} that the
appropriate description for transitions that are compatible with
inflation is the vacuum one, in terms of the tunneling of an
order parameter 
(scalar field) \cite{Linde:2005ht, Coleman:1980aw}. There is nevertheless
an amount of 
radiation, due to the collisions of bubble walls of
previous transitions.

The energy released by a transition goes partly in the acceleration of
the bubble walls, and partly in the velocity and heating of the fluid. The
features of the gravitational waves emission are different depending
on how the energy is divided in these two channels

We can understand how much of the energy released by
the transition goes into the fluid by looking at the
hydrodynamical equations for the fluid at the wall.
In fact, the fluid can reach a steady-state
configuration, at 
some time after the nucleation and at 
some distance from the wall. In that case, the transition front
(related to the bubble wall) acts as a discontinuity surface where the
energy and momentum fluxes must be conserved,
leading to the equations of conservation in the rest frame of the
wall\footnote{Here, for the 
  moment, we have 
  assumed that the interactions between 
the fluid and the wall are negligible during the bubble expansion
prior to collisions, but it is possible to extend this to
the more general case (see for example \cite{Ignatius:1993qn}).}
 \begin{align}
  {4 \ov 3} \kappa T_n^4 \, v_{(f), n} & = 
   {4 \ov 3} \kappa T_{n-1}^4 \,v_{(f), n-1}
   \label{energyflux}\\ 
  {4 \ov 3} \kappa T_n^4 \, v_{(f), n}^2 & + 
    {1 \ov 3} \kappa T_n^4 -\epsilon_n =
  {4 \ov 3}\kappa T_{n-1}^4 \, v_{(f), n-1}^2 + 
    {1 \ov 3}\kappa T_{n-1}^4 -\epsilon_{n-1}\, .
  \label{momentumflux}
 \end{align}
Here $v_{{(f)}, n (n-1)}$ is the component of the four-velocity of the
fluid in phase $n (n-1)$ locally orthogonal to the wall.

We try to find what can be the values and solutions for the velocity
that comply with inflation. Considering that
inflation allows only
small velocity and temperature perturbations,
(\ref{energyflux}), (\ref{momentumflux}) lead to 
 \beq \label{steadystate}
  {\Delta\epsilon_n \ov \kappa T^4_n} \lesssim {1 \ov 3}
    {|\delta(\kappa T_{n-1}^4)| \ov \kappa T_n^4} \, ,
 \eeq
where $\delta(\kappa T_{n-1}^4) \equiv \kappa T_{n-1}^4 - \kappa T_{n}^4$.

But from (\ref{energybound}), (\ref{Chaudhuri}), we find
 \beq \label{alphaeq}
  {\Delta\epsilon_n \ov \kappa T^4_n} > 4 {H \ov \beta_n} \, ,
 \eeq 
while\footnote{We
  estimate the enthalpy density variation compatible with inflation
using the value on superhorizon
scales as it is larger than that at sub-horizon scales. The result for
density perturbations that we
use in the formula can be rigorously obtained in the flat gauge. We
consider models of inflation, where isocurvature
perturbations are negligible.} 
 \beq \label{energydensityvariation}
  {|\delta(\kappa T_{n-1}^4)| \ov \kappa T_n^4} = 
  {|\delta \rho_f| \ov \rho_f} \approx 
  {|\delta \rho_{\text{tot}}| \ov \rho_{\text{tot}}} \sim 
   {H \ov M_{\text{Planck}}\sqrt{c_s \varepsilon}}.
 \eeq 
By looking at (\ref{sintbound}), (\ref{WMAPnormH}), we see that (\ref{alphaeq}),
(\ref{energydensityvariation}) are in 
contradiction with
(\ref{steadystate}), which must be satisfied if there exist steady state
hydrodynamical solutions 
at the wall complying with inflation. Therefore there is no such
solution.

The meaning of this result is
that, for successful transitions compatible with inflation, the vacuum energy
released in the transition goes mostly into the acceleration of the
walls, which expand at a velocity
rapidly approaching the one of light until collision. We can
therefore assume
$
  v \approx 1
$
for the wall velocity. The bubble dynamics
is well described by the approximation of bubbles in vacuum.

The gravitational waves are therefore mainly sourced
by the stress tensor of the scalar field describing the bubbles'
configuration in vacuum, with a
sub-dominant contribution from the fluid, suppressed by powers of its
small velocity.  
 
\subsubsection{Calculation of the wave spectrum}

We turn now to the detailed computation of the spectrum of gravitational
waves.
They are sourced because
of the breaking of the spherical symmetry of bubbles at collisions.
The latter are complicated events that occur out
of equilibrium. It is therefore quite difficult to give an analytic
description of their gravitational emission:
in principle we could expect a whole range of
different scales to appear in complicated
inter-correlation. 

The only way to deal with these problems is via numerical
simulations \cite{Kosowsky:1992vn, Kamionkowski:1993fg, Huber:2008hg, 
Caprini:2009fx}. On top of those, useful
analytical formulas can 
be derived \cite{Caprini:2009fx, Caprini:2007xq}.
We will obtain the spectrum of gravitational waves by extending 
the results of the numerical simulations to the case of transitions
occurring during inflation. 

The best simulations available today \cite{Kosowsky:1992vn,
  Kamionkowski:1993fg, Huber:2008hg, Caprini:2009fx} consider
short-lasting sources and 
static background (neglected expansion of the Universe, no inflation).
For nearly vacuum collisions and wall 
velocity $v \sim 1$ in their formulas, they show that
only two quantities are important  in
determining the spectrum of 
the gravitational waves: 
the overall energy density $\rho_W$ released by the transition
   and carried by the walls that collide, 
 and the time-scale $\beta_n^{-1}$ of the phase transition, which
       sets the peak frequency of emission

References \cite{Huber:2008hg, Caprini:2009fx} also show that the energy density per 
octave radiated by colliding bubbles goes like $k^3$ for small
frequencies, while for large ones it decays as $k^{-1}$. In 
particular, there is a single peak when $k \simeq \beta_n$ and the
maximum (normalized to the total energy density) is\footnote{Here, we
  specifically use the results in the more recent reference
  \cite{Huber:2008hg}. In adapting from their conventions to ours,
 one has to compare the formula we obtain using
 (\ref{hgeneralinhomogeneoussolution}) in (\ref{energyraddef}), for large $k$, with
 Weinberg's formula from \cite{WeinbergCosmologyOld}, used in
 \cite{Huber:2008hg}. Note that we define the Fourier transform of the
 energy-momentum 
 tensor as $T_{\mu\nu}(\hat k, k) = \int{d^4 k \ov (2\pi)^2}e^{ik \cdot x}T_{\mu\nu}(x, t)$,
 whereas \cite{Huber:2008hg} uses Weinberg's normalization for the Fourier
 transform. The physical equivalence of
 Weinberg's formula
and the one we obtain from (\ref{energyraddef}),
(\ref{hgeneralinhomogeneoussolution}) for large $k$ has been discussed in
\cite{Caprini:2009fx}.} 
$\sim {0.013 \ov \pi^3 M^2_{\text{Planck}}} \,{\rho_W \ov \beta^2}$. 

We stress that the transitions considered by \cite{Kosowsky:1992vn,
Kamionkowski:1993fg, Huber:2008hg, Caprini:2009fx} occur at the end or
after inflation and 
the energy carried by the walls in their cases is nearly all of the total
energy, so that $8\pi G \,\rho_W = 3 H^2$.  

We will now derive the radiated energy and spectrum of waves in our
scenario. 
There are three evident differences with the setup of
\cite{Kosowsky:1992vn, Kamionkowski:1993fg, Huber:2008hg, Caprini:2009fx}:
\begin{itemize}
 \item the Universe in our case undergoes inflation. However,
  the static approximation at the time of emission of the waves can be
  acceptable as long as their modes are $k \geq 
  aH$. This is compatible with our scenario as the typical
  scales/frequencies $\beta_n$ of the phase 
  transitions are larger than the horizon (see
  section \ref{conditionsdecayrates}).
 \item in our case, the total energy released by the walls is
   $\Delta \epsilon_{n}$ given by (\ref{energybound}), which is much
   smaller than the total vacuum energy driving inflation. 
 \item the evolution of the waves during inflation is peculiar as modes
   can exit the horizon. 
\end{itemize}
Using this information, we can now adapt the results of the
simulations and derive 
the energy density per octave radiated by colliding bubbles of the
 $n$-th transition during inflation. 
We find that initially,
at the time $t_n$ when the
sources stops being active shortly 
after the collisions (time of emission),
 \beq \label{energyperoctavecollisions}
  {1 \ov \rho_{\text{tot}}} p{d \rho^{(n)}_h(p) \ov dp}\Big|_{t=t_n} 
   \sim 
   {0.16 \ov \pi^3} \left({H \ov \beta_n}\right)^4 \, \varepsilon^2 \,
            \begin{cases}
              \left( {p_n \ov  \beta_n} \right)^3 & H_n < p_n <  \beta_n \\
              \left( {\beta_n \ov p_n} \right)   & p_n > \beta_n 
             \end{cases} \, ,
 \eeq
where $\rho_{\text{tot}}$ is the total energy density. We have
written the result in terms of the physical momentum $p_n= k
a_n^{-1}$, $a_n= a(t_n)$. 

When $k \gg \mathcal{H}$, there is a simple relation between the
spectrum (\ref{spectrumdef}) and the energy density
(\ref{energyraddef}):
 \beq \label{energyraditedandspectrum}
  k{d \rho_h(k, t) \ov dk} = k^2 {1 \ov  8 \pi a^2 G} P_h(k, t) \, .   
 \eeq
Using it, we obtain the spectrum of gravitational waves at the
time of emission $t_n$
 \beq \label{spectrumgravwavescoll}
  P_h^{(n)}\Big|_{t=t_n} = 
   {0.15 \ov \pi^2} \left({H \ov \beta_n}\right)^6 \, \varepsilon^2
         \begin{cases}
           \left( {p_n \ov  \beta_n} \right) & H_n < p_n < \beta_n \\
           \left( {\beta_n \ov p_n} \right)^3   & p_n > \beta_n 
          \end{cases} \, ,
 \eeq

We need now to evolve the spectrum after the moment of emission of the
waves. As this evolution is particular during inflation, further
differences with respect to the results in 
\cite{Kosowsky:1992vn, Kamionkowski:1993fg, Huber:2008hg,
  Caprini:2009fx} will arise. 

At the time of emission (roughly at the completion of the transition and
collisions),
the modes are within the horizon.
From (\ref{hgeneralinhomogeneoussolution}), we can see that then the
wave evolves as $\sim a^{-1}$ until the mode exits from the horizon at time
$t_{\text{ex}}$. After that, the wave remains constant until
reentering horizon. 

We need to compute the proper redshift of the modes at $t > t_n$.
For modes still within the horizon at time $t$, we find
$
{a_n \ov a(t)} = {p(t) \ov p_n}
$.
For modes that have already exit the horizon at some time
$t_{\text{ex}} < t$, instead, the redshift is only
$
 {a_n \ov a_{\text{ex}}} = {H_{\text{ex}} \ov p_n}
$.
Overall, 
the redshift factor at time $t$ for a momentum $p$ can be written as
 \beq
   {a_n \ov a(t)} 
    \equiv \chi_n {\beta_n \ov p_n} \, , \qquad 
\label{redshiftfactor}
  \chi_n  = 
  \begin{cases} 
     {H \ov \beta_n} & \text{superhorizon modes at time $t$} \quad (p(t) < H) \\
     {p \ov \beta_n} & \text{subhorizon modes at time $t$} \quad (p(t) > H) \,.
  \end{cases}
 \eeq
where we have used $H_{\text{ex}} \sim H \sim H_n$, since the
Hubble parameter does not evolve much during inflation.

Therefore, at time
$t > t_n$ (still during inflation), the spectrum generated by
the collisions of bubble of the $n$-th transition has evolved to
 \beq \label{spectrumgravwaves}
  P^{(n)}_h = 
  {0.15 \ov \pi^2} \left({H \ov \beta_n}\right)^6 \, \varepsilon^2
   \chi_n^2
         \begin{cases} 
           {f_{\beta_n} \ov p} & f_{H_n} < p < f_{\beta_n} \\
           \left( {f_{\beta_n} \ov p} \right)^5   & p > f_{\beta_n} 
          \end{cases} \, ,
 \eeq
where
 \beq \label{frequencyspectrum}
  f_{\beta_n}  = {a(t_n) \ov a(t)} \, \beta_n \, , \qquad
  f_{H_n}  = {a(t_n) \ov a(t)} \, H_n \, .
 \eeq
If we want to have a chance to detect these primordial
gravitational waves, we must consider superhorizon modes, as
subhorizon ones are more suppressed by the redshift.
Their total spectrum at time $t$ is
obtained by summing over all the phase 
transitions and is given by, using (\ref{redshiftfactor}),
(\ref{spectrumgravwaves}) for superhorizon modes,
 \be \label{totalspectrumsum}
  P_h^{\text{sup}} & = \sum_{n=1}^{N} P_h^{\text{sup}, (n)} \nonumber \\
      & = \sum_{n=1}^N \,{2 \ov \pi^2} A_n \, 
         \left({H \ov M_{\text{Planck}}}\right)^2 \,
         \begin{cases}
           {\tilde \beta_n \ov k} & a(t_n) H < k < \tilde \beta_n \\
           \left({\tilde \beta_n \ov k} \right)^5   & k > \tilde \beta_n
          \end{cases} \, , 
 \ee
where
 \beq \label{finalquant}
  \tilde \beta_n = a_n \beta_n \, , \qquad k=a_n p_n > a(t)H, \qquad
   A_n \equiv 0.07
   \left({M_{\text{Planck}} \ov H}\right)^2
   \left({H \ov \beta_n}\right)^8 \, \varepsilon^2
 \eeq
This represents the main novel result of this section.

We will deal with the later evolution of the modes, after reentering
the horizon, in section \ref{directdetectioninterferometers}.

\subsubsection{Comments on the final result}

Comparing the spectrum (\ref{totalspectrumsum}) with the one
generated in vacuum during 
inflation (see formula (\ref{vacuumspectrum}) in appendix
\ref{vacuumwaves}), we see that the 
main differences 
are as follows:
\begin{itemize}
 \item the spectrum (\ref{totalspectrumsum}) is not scale invariant
 \item due to (\ref{sintbound}), (\ref{WMAPnormH}), $A_n$ is an
   additional suppression factor,
   with respect to the waves generated by vacuum oscillations
 \item the accumulation due to a sufficiently large number of phase
   transitions $N$ could cope with the suppression and make
   the waves detectable. 
\end{itemize}
In discussing the backreaction of the waves (and possibly the
breaking down of the perturbative approach), we need to consider the
possible number $N$ of transitions.

As we see, for a few phase transitions, the waves sourced by the collisions of
bubbles are even more suppressed than those produced by vacuum quantum
oscillations, and therefore their backreaction is
negligible (perturbation theory is accurate). 

For a large number of
transitions, we have in principle a stronger backreaction. It is easily
verified, though, that the number of transitions needed to affect the
background evolution is unlikely to occur (for the range of parameters
given in section \ref{parameterssec} and $\varepsilon \sim 10^{-2}$, it
would have to be at least of the order of $N \sim 10^{12}$).

We might also wonder if the depletion of the energy density due
to gravitational 
radiation could affect the scalar density perturbations. However, from
 (\ref{totalspectrumsum}) and the above comments, it is evident that
this depletion of energy for a single transition is
very small (smaller than the linear order in perturbation
theory). Only a large number of transitions 
could have an effect (in that case, invalidating the perturbation
expansion). However, even in that case this would happen after
many transitions have occurred, that is quite late during
inflation. It would therefore affect only the late density perturbations
that are not interesting for detection at the CMBR (see also section
\ref{CMBR}). 

The spectrum
(\ref{totalspectrumsum}) is different also from the one
generated by the collisions of bubbles of first order
phase transitions that do not occur during inflation. Indeed, beside
the fact that in this second case the fluid velocity spectrum is generally an
important source and that 
 the maximum amplitude of the waves from a single transition is less
 suppressed, the most notable difference is that in our case the
 presence of modes that were superhorizon at inflation gives the
 spectrum a different dependence on the frequency (compare with
 \cite{Kosowsky:1992vn, Kamionkowski:1993fg, Huber:2008hg,
   Caprini:2009fx, Caprini:2007xq}). 
If detected, this would help in distinguishing the two cases.

Another comment concerns the sum we have performed in
(\ref{totalspectrumsum}). In fact, there is a non-zero
probability of nucleation of bubbles of phase, say, $n-2$ when the
bubbles of phase $n-1$ have not yet collided. Would this make the simple
sum rule we use inadequate?
The answer to this question is that it would not, as long as
the moments of collisions of consecutive phase transitions are
spaced out. Indeed, the 
evolution of the transition before collision (expansion of the
bubbles) does not generate gravitational waves, if the bubbles
are spherically symmetric \cite{Caprini:2007xq}.

\subsection{From turbulence, (hyper)magnetic fields,
  viscosity}\label{fluidwaves} 

The presence of a radiation dominated fluid is a consequence of the
processes of bubble collisions and 
walls decay after each phase transition. Although
sub-dominant, this
component cannot be neglected when discussing
perturbations, as it is can be even essential for having the right spectrum of
adiabatic density perturbations, for example in chain inflation
\cite{Chialva:2008zw}.  

The physics related to the plasma or hydro- dynamics could
generate gravitational waves during inflation,  
depending on the gauge coupling and the temperature
of the fluid\footnote{For an
  incomplete bibliography on the processes 
  themselves and the associated gravitational production (not during
  inflation), see \cite{MagneticPlasma}.}. This is what we want to
investigate now. 

We will make the assumption of weakly coupled plasmas, because their stress
anisotropy tensor is larger (large viscosities) than that of strongly
coupled plasmas, and therefore a potentially stronger source of
gravitational waves. Nevertheless, the anisotropy tensor cannot be
too large, otherwise it would spoil the homogeneous and isotropic
description of the Universe at large scales. It will appear that this
indeed does not happen when the parameters satisfy the bounds in section
\ref{parameterssec}.

Note also that the validity of the hydrodynamical or plasma
description depends on the relative magnitude of the Hubble scale and
the microscopical ones, such as the mean-free-path or the screening
distance. This tells us that only for some ranges of values of the
couplings the description is self-consistent. In particular, the
couplings cannot be too small, although a precise bound depends on the
model-dependent numerical factors entering the formulas of the plasma scales.

We will make use of the transport coefficients 
characterizing the transport of energy, 
momentum and charge across the fluid \cite{Arnold:2000dr}. They 
have been generally calculated within
the most well-known theories (especially QED, QCD), but usually the
results can be
extended with only minor modifications to
different higher energy theories, thanks indeed to the plasma
  or hydrodynamical approximation. 
We assume that this is our case\footnote{Our results will therefore
  apply to theories and models for which this is possible.
In the following, our
  formulas for the parameters will 
be written up to proportionality constant (indicated with a index
$_0$), which depend on the
peculiar details of the plasma under consideration (such as the number
of light versus heavy species or the rank of the gauge symmetry
group). These will not be influential for the results.}.

We are now going to analyze the various sources related to
hydro- and plasma dynamics. It will turn out that they are very much
suppressed, contrarily to what generally happens when first order
transitions occur not during inflation.
The ratio ${T \ov \beta_n}$ will be particularly important in the
following considerations. It is strongly constrained by the
requirements of efficient inflation and self-consistency of the theory
(see section \ref{parameterssec}) and this will ultimately be the reason
why these sources do
not produce a sizable spectrum of gravitational waves in our scenario. 

\subsubsection{Hydrodynamical turbulence}\label{hydroturbulence} 

Turbulence is a strong source of gravitational waves. Let us study its
occurrence in our scenario. In our case,
the generating mechanism would be the collisions of
bubbles, stirring the fluid. The typical length scale of injection is
approximately the bubble size at collisions,
$\sim \beta_n^{-1}$, and this, together with the kinematic viscosity
\cite{Arnold:2000dr} 
 \beq
  \nu = \nu_0 g^{-4}\log(g^{-1})^{-1} T^{-1} \, ,
 \eeq
yields the Reynold number
 \beq \label{reynoldnumb}
  Re = {v^{(\beta_n)}_f \beta_n^{-1} \ov \nu}
     \sim v^{(\beta_n)}_f {T \ov \beta_n }g^{4}\log(g^{-1}) \, ,
 \eeq
where $v^{(\beta_n)}_f$ is the characteristic velocity of the fluid flow at the
injection scale.

Large Reynold numbers signal the onset of
turbulence. We therefore see 
from (\ref{reynoldnumb}) that the relevant condition is
 \beq
  {T \ov \beta_n} = 
    \begin{cases}
      \gg 1 & \text{turbulence} \\
      \lesssim 1 & \text{no turbulence} .
    \end{cases}
 \eeq 
Looking at
(\ref{WMAPnormH}, \ref{bathtemperature}, \ref{sintbound}), we find
that in our scenario 
 \beq
  {T \ov \beta_n} \, < \, 4 \, . 
 \eeq
Therefore, turbulence does not occur and a sizable emission of
gravitational waves 
is not possible. The reasons for this are the smallness of the scales of
injection of 
energy and the rapidity of the bubble evolution, which are consequences of
the requirements of small backreaction on inflation and self-consistency of the
theory (section \ref{parameterssec}).

\subsubsection{Plasma physics and gauge fields} 

The radiation fluid is likely
charged (therefore it is generically a plasma) and 
local charge asymmetry and currents can be
generated\footnote{For example, it is 
enough to have different mean free paths (different couplings)
for the various species to generate a local charge
asymmetry, in
presence of particle-antiparticle number asymmetry. Then,
 the collisions among bubbles impart a vorticity
to the charged fluid, with the creation of microscopic currents.
The number asymmetry can occur because
bubble collisions and first order transitions are likely to provide
the right conditions \cite{Sakharov:1967dj}. Nevertheless,  
a precise statement can be made only with a detailed
model
\cite{MagneticPlasma}.}.  
We investigate now whether non-zero long-range gauge
fields 
generated by these currents could 
represent a sizable source of gravitational waves during inflation.

The only long-range
fields that can survive in a plasma for enough time are magnetic
fields associated 
with a $U(1)$ gauge symmetry (hypermagnetic fields), see section 5.6 in
\cite{Giovannini:2003yn}.
The precise form of the generated fields depends on the
details of the 
bubble growth, the couplings, the tunneling events, the non-equilibrium
dynamics. 
Nevertheless, it is possible to draw sufficient conclusions
from general considerations.

The fields are generated by the currents at
the scale $L_p \sim \beta_n^{-1}$ of the bubbles\footnote{Due to the
  different orientation of field lines at the bubble scale, it is also
  possible to assume that the field spectrum is stochastic, in this
  way preserving the global isotropy of spacetime.}
\cite{MagneticPlasma} and then possibly enhanced by magnetic
turbulence, which is signaled by a large magnetic Reynold
number. This is defined as  
  \beq
   Re_\mu = {v_f^{(\beta_n)} L_p \sigma_c \ov 4 \pi} \, ,
  \eeq
  where $L_p$ is the typical scale at which magnetic fields are
  generated and $\sigma_c = \sigma_0 {T \ov  g^{2} \log(g^{-1})}$ is
the (hyper)electrical conductivity at high temperature.
 
In our case,
 \beq
  Re_\mu \sim   v_f^{(\beta_n)}{T \ov \beta_n g^{2} \log(g^{-1})} \, ,
 \eeq
which, even for reasonably small couplings ($g \sim 0.1$) is much less
than 100, given the bounds in section \ref{parameterssec}.
Therefore there is no hypermagnetic turbulence
and it is unlikely that the small scale hypermagnetic fields
appreciably source gravitational waves\footnote{In principle,
  there is still a residual but unlikely possibility that in specific
  models even 
  the small scale hypermagnetic fields can be strong enough to
  generate a sufficiently sizable amount of gravitational
  emission. We
  will not discuss this case.}. 

\subsubsection{Stress tensor from fluid viscosity}

The form of the anisotropic stress tensor in general
relativity for a fluid with velocity
$\vec u$ is \cite{WeinbergCosmologyOld}
 \beq \label{viscousstresstensor}
  \pi^{(\text{visc})}_{ij} = -\varsigma 
  \left(\partial_j u^{(f)}_i+ \partial_i u^{(f)}_j-{2 \ov 3}
  \nabla\cdot \vec u^{(f)}\delta_{ij} \right) 
   - \zeta \nabla\cdot \vec u^{(f)}\delta_{ij} . 
 \eeq 
$\zeta$ is the bulk viscosity. For a relativistic fluid, $\zeta$ is
 vanishing, as it is given by  
    \beq \label{bulkviscosity}
     \zeta = \zeta_0 T^3 g^{-4} 
   \left({1 \ov 3} - {\partial \rho_f \ov \partial P_f}\Big|_{n_f} \right) \, ,
    \eeq
where $\rho_f, P_f, n_f$ are respectively the energy, 
pressure and number density.

The part of $\pi^{(\text{visc})}_{ij}$ proportional to 
the shear viscosity $\varsigma$ could instead be important, as
    \beq \label{shearviscosity}
     \varsigma = \varsigma_0 {T^3 \ov g^{4} \log(g^{-1}) } \, 
    \eeq
for a weakly coupled fluid.
Nevertheless, it turns out that this source is negligible as well.

In fact, the  energy per octave radiated by an anisotropic stress
tensor $\Pi_{i j}$, normalized by the total energy density, at
emission time $\eta= \eta_*$ on a scale $k > \mathcal{H}$ is 
 \beq
  \Omega_* \equiv 
  {1 \ov \rho_{\text{tot}}}k {d \rho_h(k) \ov dk}\Big|_{\eta=\eta_*} =
  {16 \ov \rho_{\text{tot}}}
   G k^3 a_*^2 \int d\Omega
   \langle \Pi_{i j}(k, k) \Pi^*_{i j}(k, k) \rangle  \, , 
 \eeq 
where $a_*= a(\eta_*)$, while $\Pi_{i j}(k, k) = 
\int {d\eta\, d^3x \ov (2\pi)^{2}}
e^{i k \eta-i \vec k \cdot \vec x} \Pi_{i, j}(\vec x, \eta)$ 
and $\int d\Omega$ is the angular integral.

In the case of (\ref{viscousstresstensor}), considering only the shear
viscosity part, we have
 \beq
  \Omega^{(\text{visc})} \simeq  {16 G \varsigma^2 \ov \rho_{\text{tot}}} 
  \mathcal{V}^2 \,    
 \eeq
where $\mathcal{V}^2 \equiv a_*^2 k^5 \lambda^5 v_{\partial,\,\lambda}^2$
and $v_{\partial,\,\lambda}^2(\lambda k,k)$ is
the dimensionless square mean-field gradient of the
velocity (integrated over the angles and Fourier transformed in the
time dependence)
obtained by smoothing
over the comoving length scale
$\lambda$ and in which we have also absorbed the dependence on $k\lambda$
from the velocity spectrum. The scales
suitable for hydrodynamical treatment must be larger than the
coherence scale of the velocity, therefore in our scenario
$\lambda \sim \tilde \beta_n^{-1}$.

On the other hand, at the moment of emission, the energy per octave
radiated by the 
collision of bubbles of one phase transition is
(\ref{energyperoctavecollisions}), here written 
for comoving wavenumbers, 
 \beq
  \Omega^{(\text{coll})} = 
   {0.16 \ov \pi^3} \left({H \ov \beta_n}\right)^4\varepsilon^2 
   \left({\tilde \beta_n \ov k}\right)^{-3} \, .
 \eeq 
We have considered wavenumbers
appropriate for the comparison with the scales of the hydrodynamical
description, which means $k < \tilde \beta_n$.

For scales $k \lesssim \lambda_f^{-1}$, where
$ \lambda_f = \lambda_0 g^{-4}\log(g^{-1})^{-1} T^{-1}$ is the mean-free-path, we therefore find
 \beq
  {\Omega^{(\text{visc})} \ov \Omega^{(\text{coll})}} \lesssim
   v_{\partial,\,\lambda}^2\big |_{k=\lambda_f^{-1}}
  \ll 1 \, ,
 \eeq
where we have used (\ref{sintbound}, \ref{WMAPnormH},
\ref{bathtemperature}) and the final estimate takes into account that 
$v_{\partial,\,\lambda}^2$ is a small perturbation during inflation.

The energy radiated in gravitational waves from
viscosity is therefore smaller than the one coming from bubble
collisions. 

\subsubsection{Thermal fluctuations}

There are also other possibilities for generating (hyper)magnetic 
fields in a plasma, beside local currents from bubble dynamics:
for example quantum and thermal fluctuations. In both cases
the generated fields can source gravitational waves.

A complete analysis of these sources would go beyond the scope of this
paper, as they are not strictly present only when
first order transitions occur and the results
would depend on specific details of the models. 

In fact, quantum fluctuations can generate
gauge fields extended at appreciable scales only when their
couplings to the metric are not conformal symmetric (see for example
\cite{Turner:1987bw} and reference therein). On the other hand, the
excitation of gauge fields via thermal fluctuations has been studied
in flat Minkowski space via
thermal field theory or the Boltzmann 
equation \cite{MagneticfieldsThermal}, and the
extension of those results to a rapidly expanding
Universe is not straightforward.

\section{Detection}\label{discussionconclusion}

Our analysis indicates that the dominant
contribution to the spectrum of gravitational waves from
first order phase transitions during inflation, is sourced
by collisions of bubbles and has the spectrum (\ref{totalspectrumsum}). 

We will now discuss the possibility of detection of this spectrum both
in the CMBR anisotropies and by direct measurement at
interferometers. Although
  detailed, our analysis will not be precise down to numerical factors
  of order one, which depend on the particular high-energy models.

\subsection{CMBR}\label{CMBR}

Primordial gravitational waves affect both the temperature
anisotropies and the polarization of the CMBR. Experiments such as
CMBPol and Planck will investigate these observables in the next future.
Here, we will concentrate on the temperature
anisotropies, considering the so-called
tensor (T)-to-scalar (S) ratio
 \beq \label{tensorscalarratio}
  r= {C^T_\ell \ov C^S_\ell} \, ,
 \eeq
where $C^T_\ell, \,C^S_\ell$ come from the decomposition of the
spectrum of temperature 
anisotropies in two directions $\mathbf{l_1}, \mathbf{l_2}$ with
$\mathbf{l_1}\cdot \mathbf{l_2} = \cos(\theta)$,
 \beq \label{multipolestemperatureanisotropies}
  \left\langle {\delta T \ov T}(\mathbf{l_1}){\delta T \ov T}(\mathbf{l_2})
   \right\rangle
   = {1 \ov 4\pi} \sum_{\ell=2}^\infty (2\ell+1)(C^T_\ell+C^S_\ell)
   P_\ell(\cos\theta) \, .
 \eeq

The precise computation of $r$ would require a 
numerical evaluation of (\ref{multipolestemperatureanisotropies}) using
the spectrum (\ref{totalspectrumsum}). However, for our purposes it suffices
an approximate analytical computation. We follow
\cite{Mukhanov:2005sc}, assuming $\ell \gg 1$\footnote{Recall that the 
multipoles we can consider are such that $\ell < 200$.}. 

Using $k= {\ell H_0 \ov 2}, \, a_0=1$, we obtain
 \beq
  \ell (\ell+1) \, C^T_\ell \, \sim {5 \pi^2 \ov 64} \,\,
    P_h^{\text{sup}}({\ell H_0 \ov 2})
 \eeq
and therefore, for the spectrum (\ref{totalspectrumsum}) and 
$\ell (\ell+1) \,C^S_\ell={4 \ov 25} 
  {H^2 \ov 4\pi c_s \varepsilon M_{\text{Planck}}^2}$,
 \beq \label{ratiotensorscalar}
  r \sim 12.3\, c_s\, \varepsilon 
    \sum_n^N\, A_n\,{\ell_{\tilde \beta_n} \ov \ell}, 
      \qquad \ell_{\tilde \beta_n}= {2 \tilde \beta_n \ov H_0} \, ,
 \eeq
where we have considered only wavenumbers in the range 
$ a_n H < k < \tilde \beta_n$ since the contribution from larger wavenumbers is
very suppressed (see (\ref{totalspectrumsum})). 

Equation (\ref{ratiotensorscalar}) is very different from the result
for tensor modes generated by vacuum
fluctuations in single-field slow-roll/chaotic models, which would be
$r=8\varepsilon$: indeed, (\ref{ratiotensorscalar}) shows
{\it i}) a dependence on the multipole $\ell$ due to  the lack of scale
invariance in 
(\ref{totalspectrumsum}), {\it ii}) an additional suppression factor
$A_n$ (\ref{finalquant}), and {\it iii}) a possible accumulation due to
many phase transitions. 

We would like to understand how many transitions are necessary
to have a detectable amount of tensor modes (we require $r \sim 0.07$).
To proceed, we will make a simplifying
assumption: a negligible dependence on $n$ for the ratio ${\beta_n
\ov H_n}$. We will also take $\varepsilon \sim 10^{-2}$
    as a typical value, for illustrative purposes.

We also limit our estimate to the multipoles corresponding to the frequencies
$p = \beta_n$, which is
the typical one of the collisions, and
$p \simeq H_n$, which is the lower limit of validity for the numerical
simulations we have extended. We indicate these multipoles in
the formulas as $\ell_* = \{\ell_{\tilde \beta_n},  \ell_{\mathcal{H}_n}\}$.
For these specific frequencies, also because of the simplifying
assumption we have made above, the ratio 
${\ell_{\tilde \beta_n} \ov \ell_*}$ is independent of $n$. 

We define
 \beq \label{suppressionfactor} 
   \sum_{n=1}^{N} A_n {\ell_{\tilde \beta} \ov \ell_*} \,\, \equiv \,
   A \, {\ell_{\tilde \beta} \ov \ell_*} \,\,
   \simeq \, 0.07 \, N 10^{8} \varepsilon^{-1}
   \left({H \ov \beta_n}\right)^8 \, \varepsilon^2
   {\ell_{\tilde \beta} \ov \ell_*}\, ,
 \eeq
where we have used (\ref{WMAPnormH}), (\ref{finalquant})
and $\ell_{\tilde \beta}= {2 \tilde \beta \ov H_0}$.

Taking into account (\ref{sintbound}),
$A$ lies in the range
 \beq \label{Arange}
    1.1 \times 10^{-6} \, N \, \varepsilon \, 
    \left({S_E \ov \pi^2}\right)^{{8 \ov 5}} \,
    \lesssim \,  A  \, \lesssim 0.07 N \, \varepsilon \, 
 \eeq
and therefore the scalar to tensor ratio (\ref{ratiotensorscalar})
falls within the interval
 \beq
  1.5 c_s \,10^{-5} \, N \, \varepsilon^2 \, 
    \left({S_E \ov \pi^2}\right)^{{8 \ov 5}} \,
     {\ell_{\tilde \beta} \ov \ell_{*}} 
   \leq \, r \, \leq
  0.9 c_s N \, \varepsilon^{2} \,
    {\ell_{\tilde \beta} \ov \ell_{*}} \, .
 \eeq

If $k=\tilde \beta$, we see that 
$N \sim 7.7 \times 10^2$
transitions with the minimal allowed transition rate (${\beta \ov H} \sim 10$)
would be sufficient to have  $r\sim 0.07$. But the
number of transitions necessary for detection rapidly increases for
faster rates, up to $N \sim 3 \times 10^{8}$ for the fastest rate in
(\ref{sintbound}).

If, instead, it is $k = a_n H_n$, we find that 
$N \gtrsim 77$ transitions with slow 
rate (${\beta \ov H}\sim 10$) would be 
detectable via the CMBR, but for the fastest transitions we
need at least $N \sim 6 \times 10^6$ of them.

It appears therefore that a not too large number of transitions occurring
at a slower rate could 
be detectable, especially for lower multipoles. 
However, as the rate of the
transitions increases, the spectrum rapidly 
decreases and becomes undetectable. 

Note that in many of the studied models of transitions
outside inflation it is found that the ratio
${\beta \ov H}$ is larger than ${\beta \ov H}\sim 10$. However, the
situation for first order transitions in models 
and theories valid at the scales of inflation could be 
different. 
Because of our limited 
knowledge of the relevant theories at very high energy, such as the
string landscape, we cannot discard those values of ${\beta \ov H}$.
It is therefore important to analyze this point in future
research within concrete models of interest.

The suppression of the emission from
fast transitions also occur for transitions not 
during inflation, but in the inflationary scenario the
phenomenon is much more pronounced as the suppression factor goes like
$\left({H \ov \beta}\right)^8 \varepsilon^2$ for each transition,
see (\ref{totalspectrumsum}).

\subsection{Direct detection}\label{directdetectioninterferometers}

We will investigate the direct detectability of the gravitational
signal by interferometers considering the cases of LIGO, LISA and
DECIGO\footnote{We will actually consider Ultimate DECIGO.}.
To do so, we need to evolve the
spectrum after inflation ends, from the moment when the mode
reenters the horizon until now. 

Since, differently from the
previous sections, we are now
evolving the waves also after inflation, we will indicate as 
$H_{\text{infl}}$ the Hubble rate during inflation, to avoid
any possible confusion.

From (\ref{highfreq}), we see that after reentering the
horizon, $h$ evolves as $\sim a^{-1}$. We define the transfer
function $\mathcal{T}(p) = {a_p \ov a_0}$, where $a_p$ is the scale factor
at the time of reentering for the physical momentum $p$
measured today.
As discussed in \cite{Smith:2005mm}, the sensitivities of LIGO, LISA, and
DECIGO peak around frequencies which had to be within the horizon well
before matter-radiation equality and nucleosynthesis. Therefore, assuming
adiabatic expansion after the end of inflation, 
\cite{Smith:2005mm} 
 \beq \label{transfunct}
  \mathcal{T}(p) = 2.1 \times 10^{-20} \left(0.63 \text{Hz} \ov p\right)
    \left({100 \ov \kappa_p}\right)^{{1 \ov 6}}
 \eeq
where $\kappa_p$ is the number of effective relativistic degrees of
freedom contributing to the energy density at the moment of the
reentry of the scale $p^{-1}$.

We discuss the possible detection of the primordial
waves in terms of the {\em strain amplitude}\footnote{Obtained re-writing formula
  (3.2) in \cite{Maggiore:1999vm} with our
conventions in (\ref{spectrumdef}), (\ref{energyraddef}). See also section 2.2 in \cite{Maggiore:1999vm}.}
 \beq
  \bar h_p = \mathcal{T}(p) \sqrt{{\pi P^{\text{super}}_h(p) \ov 2 p}} \, .
 \eeq
By using (\ref{transfunct}) and (\ref{totalspectrumsum}), we find
 \be \label{strainform}
  \bar h_p = 2.1 \times 10^{-20} \left({0.63 \text{Hz} \ov p}\right)
    \left({100 \ov \kappa_{p}}\right)^{{1 \ov 6}}
    \sum_n {H_{\text{infl}} \ov \sqrt{\pi} M_{\text{Planck}}}
     \sqrt{{A_n \ov  p}} \, \sqrt{{f_{\beta_{n}, 0} \ov p}}
 \ee
where we have considered only wave-numbers
$ a_n H_{\text{infl}} < k < \tilde \beta_n$, since the contribution
from larger ones is very suppressed, and we sum over the different
transitions.  

Here,
 \beq \label{betanow}
  f_{\beta_{n}, 0} = {a_n \beta_n \ov a_0} = \beta_n e^{-\mathcal{N}_n} 
   8.0 \times 10^{-14}  
    \left({100 \ov \kappa_{\text{end}}}\right)^{{1 \ov 3}}
    {1 \text{GeV} \ov T_{\text{end}}} \, ,
 \eeq
$\mathcal{N}_n$ is the number of e-foldings from the moment of
collision of bubbles of the $n \to n-1$ transition until the end of
inflation, and 
$T_{\text{end}}, \kappa_{\text{end}}$ are evaluated at the end
of inflation. 
In the following we will assume 
 $
  \kappa_n, \kappa_p, \kappa_{\text{end}} \approx 10^2 \,
 $ for illustrative reasons. These are also the typical values
  in GUTs and minimal supersymmetric models.

To evaluate (\ref{strainform}, \ref{betanow}), we need to compute
the temperature $T_{\text{end}}$, which is given by
 \beq \label{fintemp}
  T_{\text{end}} = \sum_{n=1}^{N} T_n e^{-\mathcal{N}_n} + T_f \, .
 \eeq 
Here $T_n \sim T$ is the temperature of a single phase transition, see
(\ref{bathtemperature}), and $T_f$ comes from a possible final decay
of the inflaton (reheating), which could be
also one last phase transition with large backreaction. 

We can perform the sum in (\ref{fintemp}) by considering that the last
transition occurred $\mathcal{N}_{\text{last}}$ e-foldings before the
end of inflation and that the earlier transitions and collisions times 
were spaced out by intervals of
$\Delta \mathcal{N}_n \approx \Delta \mathcal{N}$
e-foldings\footnote{Here, we have made the simplifying assumption that
the intervals of e-foldings are approximately the same for all
transition. The crudeness
of such an approximation can be determined only by building
detailed models, but this goes beyond the scope
of this paper.}. Then we define
 \beq \label{totefoldtrans}
  \sum_{n=1}^{N} e^{-\mathcal{N}_n} = e^{-\mathcal{N}_{\text{last}}} 
    {1 - e^{- N \Delta \mathcal{N}} \ov 1 - e^{- \Delta \mathcal{N}}} 
    \equiv F(N, \Delta\mathcal{N}, \mathcal{N}_{\text{last}})\, .
 \eeq
Assuming finally, for simplicity, that $T_f  < \sum_n T_n
e^{-\mathcal{N}_n}$, we obtain 
 \beq \label{Tendapprox}
  T_{\text{end}} \approx T \, F(N, \Delta\mathcal{N}, \mathcal{N}_{\text{last}})
   \, .
 \eeq

Let us recall now that (\ref{totalspectrumsum}) is
valid only for $p_n \geq   H_{n}$. These high frequencies could fall
within the range of sensitivity of the interferometers 
only after sufficient redshift. The sum over transitions in
(\ref{strainform}) will therefore start from the transition
$n_{\text{min}}$ for which at least the smallest frequency that
we can consider at the time of production $t_{n_{\text{min}}}$ has had 
the necessary redshift until now.
That frequency is $p^{H_{n_{\text{min}}}}(t_{n_{\text{min}}}) =
H_{n_{\text{min}}} \sim H_{\text{infl}}$, which is redshifted today to
 \beq \label{pHnow}
  p^{H_{n_{\text{min}}}}(t_0) = 
    {a_{n_{\text{min}}} \ov a_0}\, p^{H_{n_{\text{min}}}}(t_{n_{\text{min}}}) =
   1.2 \times 10^{{19 \ov 2}}
   {e^{-\mathcal{N}_{n_{\text{min}}}} \ov F(N, \Delta\mathcal{N}, \mathcal{N}_{\text{last}})} 
   \, \text{Hz}
 \eeq

\paragraph{Comparison with experimental setups}
~~

We are now ready to compare our results with the sensitivities of
LIGO, LISA and Ultimate DECIGO. The latter peak around certain frequencies as
follows \cite{Maggiore:1999vm, Takahashi:2004yr}:
 \be 
  &\text{LIGO}: && \bar h_f  \sim 10^{-23} \, \text{Hz}^{-{1\ov 2}} 
     && \text{at} \qquad f  \sim 100 \, \text{Hz} \\
  &\text{LISA}: && \bar h_f  \sim 4 \times 10^{-21} \, \text{Hz}^{-{1\ov 2}} 
     && \text{at} \qquad f   \sim 10^{-3} \, \text{Hz} \\
  &\text{UDECIGO}: && \bar h_f  \sim 10^{-27} \, \text{Hz}^{-{1\ov 2}} 
     && \text{at} \qquad f  \sim 0.1 \, \text{Hz} \, .
 \ee

The results we list in table \ref{directdetecttable} have been
obtained by asking for the strain amplitude (\ref{strainform}) to be 
within the
sensitivity of the detectors when evaluated at the respective peak
frequencies. The strain is computed using (\ref{strainform}, \ref{betanow},
\ref{Tendapprox}, \ref{pHnow}).
We have chosen $S_E^c \approx O(1), \, \varepsilon \sim 10^{-2}$ 
as indicative values and assumed for ${\beta_n \ov H_n}$ a
negligible dependence on $n$. The range of values considered for $A =
\sum_n A_n$ is given by (\ref{Arange}).
 
The detectability of the emitted waves depends on
the number of transitions, their timescales and
the number of e-foldings that must have been occurred to sufficiently
redshift their frequencies after the waves were emitted. These quantities enter
the strain through the function 
$F(N\!\!-\!\!n_{\text{min}},{\Delta\mathcal{N} \ov 2}, 0)$ (see
(\ref{totefoldtrans})). Table  
 \ref{directdetecttable} shows the detectability bounds on this
 function for given $N, n_{\text{min}}, \Delta\mathcal{N}$.

\begin{table}[t]
\vspace{0.2cm}
\centering
\renewcommand{\arraystretch}{2}
\begin{tabular*}{1.022\textwidth}{@{\extracolsep{\fill}}||c|c|c|c||}
\hline  
  & $f$ \,(Hz) & \multicolumn{2}{c||}{Detectability} \\ \hline
  &   &  No if & Yes for all possible ${\beta \ov H}$ if \\
\hline \hline 
\!LIGO\! & $100$ &  
   $\!\!
\text{{\small $F(N\!\!-\!\!n_{\text{min}}, {\Delta\mathcal{N} \ov 2}, 0)$}}
    \,<\,  1.6 \times 10^{6}$ & 
   $\!\!
\text{{\small $F(N\!\!-\!\!n_{\text{min}}, {\Delta\mathcal{N} \ov 2}, 0)$}}
    \,\geq\, 4.4 \times 10^{8}$    \\ \hline
\!LISA\! & $10^{-3}$ & 
   $\!\!
\text{{\small $F(N\!\!-\!\!n_{\text{min}}, {\Delta\mathcal{N} \ov 2}, 0)$}}
    \,<\,  19.7$ &
   $\!\!
\text{{\small $F(N\!\!-\!\!n_{\text{min}}, {\Delta\mathcal{N} \ov 2}, 0)$}}
    \,\geq\,  5.5 \times 10^{3}$  \\ \hline
\!UDECIGO\!  & $1$ & 
   $\!\!
\text{{\small $F(N\!\!-\!\!n_{\text{min}}, {\Delta\mathcal{N} \ov 2}, 0)$}}
    \,<\,   4.9 \times 10^{-3}$ &
   $\!\!
\text{{\small $F(N\!\!-\!\!n_{\text{min}}, {\Delta\mathcal{N} \ov 2}, 0)$}}
    \,\geq\, 1.4 $  \\ \hline
\end{tabular*}
\caption{Bounds on $N, n_{\text{min}}, \Delta\mathcal{N}$ via the
  function $F$ for the direct detection of primordial gravitational
    waves from phase transitions during inflation.}
\label{directdetecttable}
\end{table}

To understand better what these results mean for the physical
parameters, we specialize to the case of chain inflation
\cite{Chialva:2008zw} as an illustrative example.  
This one is a fast tunneling model, where inflation stops shortly after
one last phase 
 transition (so that $\mathcal{N}_{\text{last}} \sim 0$) and where
 $N \gg 1$ and $\Delta \mathcal{N} \sim {H_{\text{infl}} \ov \beta}$. 
From (\ref{totefoldtrans}), in this case
 \beq \label{chaininflationefoldings}
  F(N-n_{\text{min}}, \text{{\small ${\Delta\mathcal{N} \ov 2}$}}, 0) \sim
  \begin{cases}
   2{\beta \ov  H_{\text{infl}}} & 
    \text{for $N-n_{\text{min}} \gg {\beta \ov H_{\text{infl}}}, 
               \quad {\beta \ov H_{\text{infl}}} \gg 1$} \\
   N-n_{\text{min}} & 
    \text{for ${\beta \ov H_{\text{infl}}} \gg N-n_{\text{min}}, \quad 
       {\beta \ov H_{\text{infl}}} \gg 1$}  
  \end{cases} \, .
 \eeq

Using (\ref{chaininflationefoldings}), (\ref{sintbound}) and table
\ref{directdetecttable}, 
we see that, for instance, in the case 
$N-n_{\text{min}} \gg {\beta \ov H_{\text{infl}}}$ (many transitions), the 
gravitational waves produced by bubble collisions in chain inflation
will be 
\begin{itemize}
 \item detectable at Ultimate DECIGO for all allowed values of
         ${\beta \ov H_{\text{infl}}}$
 \item detectable at LISA for 
  ${\beta \ov H_{\text{infl}}} < 11$. 
\end{itemize}
In particular, if we take the most favorable case\footnote{Recall the
comments at the end of section \ref{CMBR}.}, 
${\beta \ov H_{\text{infl}}} \approx 10$, the detected waves 
would have been emitted by phase transitions occurring at least
20 e-foldings before the end of inflation for Ultimate DECIGO, or 27 for LISA.
Unfortunately, the gravitational waves would be undetectable at LIGO. 

If detected, these waves could be distinguished from those of
transitions occurring outside inflation, thanks to the different frequency
dependence of the spectrum of modes that were superhorizon at
inflation\footnote{Note however that the strain
  amplitude for these waves decays more rapidly with the frequency: as
  $p^{-2}$ for frequencies smaller than the 
redshifted scale of the transition and  as $p^{-4}$ for larger
frequencies. The signal from transitions not 
  during inflation goes instead like $p^{-1}$ and $p^{-3}$
  respectively in the two ranges of frequencies.}.

\subsection{Nucleosynthesis bound}

There is an important constraint on gravitational
emission: the waves that reentered the horizon before nucleosynthesis
must not interfere with it, and therefore satisfy 
\cite{Maggiore:1999vm}
 \beq \label{strainnuclbound}
  \bar h_p < 1.99 \times 10^{-21} {1 \text{Hz} \ov p^{{3 \ov 2}}} \, .
 \eeq
We study this constraint just in the case of chain inflation, as an
example. From formulas
(\ref{strainform}), (\ref{chaininflationefoldings}), we see that in that case
(\ref{strainnuclbound}) implies
 \beq \label{snuclconstr}
  {\beta \ov H_{\text{infl}}} \gtrsim 0.21 \quad \forall \, p,
 \eeq
which is certainly satisfied by the range
(\ref{sintbound}) of allowed values for ${\beta \ov H_{\text{infl}}}$.
We conclude that the bound from nucleosynthesis does not rule out
chain inflation.

\section{Conclusions}\label{discussion}

In this work we have studied the production, features and
detectability of gravitational waves in models of the early Universe
where first order phase transitions occur during inflation.

We have described these scenarios via some physical parameters,
whose values have been constrained and bounded by an analysis 
of the self-consistency of the theory (in particular efficient inflation
taking place, 
homogeneity and isotropy at large scales).

The emission and features of gravitational waves are strongly affected
by these bounds and by the specific dynamics during inflation
(such as the exit from horizon of the modes). The resulting spectrum
is different from the one due to vacuum oscillations or
first order phase transitions occurring not during inflation. 

The first important feature is that the waves from a single transition
during inflation are very much suppressed but the accumulation due to
many transitions could make them sizable.
Second, turbulence and (hyper)magnetic
fields are a negligible source of waves, contrary to what generally
happens when the transitions occur not during inflation. The collisions of the
bubbles at the end of the transitions represent the
prominent source of waves, yielding a non-scale-invariant spectrum
with different frequency dependence for modes that exit horizon during
inflation, compared to the spectrum sourced by transitions not
occurring during inflation.

We have also studied the experimental detectability of the waves.
The main points emerging from this part of the analysis are that: 
 \begin{itemize}
  \item a not too large number of slow (but still successful) transitions
     occurring during the inflationary era could leave observable
     marks in the CMBR 
     anisotropies (large tensor-to-scalar ratio for reasonable values 
    of the parameters). As the fastness of
   the transitions increases, the signal rapidly weakens, requiring
   accumulation from a 
   large number of them to be measurable,
  \item direct detection via interferometers could be possible at
   LISA and Utimate DECIGO for
    modes that were superhorizon at inflation and for a large number
    of transitions. However, at LISA the detection could occur only
    for the most optimistic scenario (slow transitions). 

   We could distinguish 
   transitions during inflation from those outside inflation thanks to
   the different frequency dependence of the spectra, 
  \item the nucleosynthesis bound is easily satisfied by the models, in
    particular chain inflation is not ruled out. 
 \end{itemize}

\appendix

\section{Appendices}

\subsection{Friedman and Chaudhuri equations}\label{basic}

With a Friedman-Robertson-Walker ansatz 
 \beq
  ds^2 =  -dt^2 + a(t)^2d\vec x^2
 \eeq 
for the background metric, the
Friedman and Chaudhuri equations read
 \be 
  H^2 & = {8\pi G \ov 3} \, \sum_\ell \rho_\ell = 
    {8\pi G \ov 3} \, \rho_{\text{tot}} \label{Friedman}\\
  \dot H & = -4\pi G  \, \sum_\ell (\rho_\ell +P_\ell) \, , \label{Chaudhuri}
 \ee
where $\rho_\ell, P_\ell$ are respectively the energy and pressure
density of the component $\ell$ of the Universe (in our case scalar
fields, radiation). The sum is over all components. Note that during
inflation $\rho_{\text{tot}} \sim \rho_{\text{vacuum}}$. 

The conformal time is related to the cosmic one by 
$d\eta = {dt \ov a(t)}$. 

\subsection{Timescale of transitions}\label{timescaletrans}

We call $p_n(t)$ the {\em vacuum persistence probability}, that is
the probability for a point in the Universe to remain in the $n$-th
vacuum at time $t$. By neglecting the possibility to tunnel
directly to distant phases, it obeys the most general equation
 \beq \label{equationphasepersistence}
  \dot p_n = -\widetilde \Gamma_n \, p_n + \widetilde \Gamma_{n+1} \, p_{n+1}
 \eeq

The conclusion of the phase transition is indicated by several markers
\cite{Chialva:2008zw, Turner:1992tz}:  {\it i}) the time $t_{c, n}$, when the
patch of the Universe occupied by the old phase starts to contract, {\it ii})
the time $t_{p, n}$, when percolation occurs, {\it iii}) the time $t_{s, n}$,
when  the probability $p_n(t)$ for a point to remain in phase $n$ at
time $t$ has dropped below a suitable small number.
   
This last requirement is actually not a faithful  signal of the
completion of the phase transition (recall for example the issues in
Old Inflation), but it yields an indication that is generally in
accordance with the other two more significant conditions, when they
occur, beside being physically reasonable.  For a rapid transition it
can be shown that $t_{s, n} \gtrsim t_{c, n}, t_{p, n}$. 
   
The time-scale of the transition can be defined as \cite{Turner:1992tz}
 \beq
  \beta_n^{-1} \sim t_{s, n} - t_{i, n}
 \eeq
where $t_{s, n}, t_{i, n}$ are such that\footnote{Note that also the
  earlier and  successive 
  phases are sub-dominant at $t_{i, n}$, but in order to  estimate
  $\beta_n^{-1}$ we need to consider only the phases $n$ and $n+1$.}
 \be
   p_n(t_{s, n}) & = e^{-M} \ll 1 & p_{n+1}(t_{s, n}) & = e^{-M-q} \ll 1 \\ 
   p_n(t_{i, n}) & = e^{-m} \sim 1 & p_{n+1}(t_{i, n}) & = e^{-M+q'} \ll 1 \, 
 \ee 
for suitable $M, M-q, m+q' \gg 1, \quad m < 1$.

If we have $\widetilde
\Gamma_n > H$ from the onset, and therefore the phase transitions are
occurring very rapidly, we can expand
    \beq
     p_n(t_{s, n}) \sim p_n(t_{i, n}) + \dot p_n(t_{i, n}) (t_{s, n} - t_{i, n})
    \eeq
and from (\ref{equationphasepersistence}), we obtain 
    \be
     p_n(t_{s, n})-p_n(t_{i, n}) & =  e^{-M}-e^{-m} \\ 
           & \Downarrow & \nonumber \\
     \left(-\widetilde \Gamma_n (t_{s, n} - t_{i, n}) +1 \right) e^{-m} & = 
     \left(-\widetilde \Gamma_{n+1} e^{q'}(t_{s, n} - t_{i, n}) +1 \right) e^{-M} 
     \, .
    \ee 
Since $e^{-M} \ll 1$, 
    \beq \label{betafasttunnelling}
     t_{s, n} - t_{i, n} \approx \widetilde \Gamma_n^{-1} \equiv \beta_n^{-1} \, .
    \eeq

For tunneling rates depending on time, expanding the tunneling
action/free energy around $t_{s, n}$ as 
$S^{(n)}_E(t) \simeq  S^{(n)}_E(t_{s, n}) - \beta_n \, (t-t_{s, n})$,
one can also find for the decay rate per unit time and volume
\cite{Turner:1992tz} 
 \beq \label{betaevolvingtunnelling}
  \Gamma_n = C e^{-S^{(n)}_E(t)} \qquad
  \beta_n = -{d S^{(n)}_E \ov d t}\Big|_{t_{s, n}} \, .
 \eeq
As we see from (\ref{betafasttunnelling},
\ref{betaevolvingtunnelling}), $\beta_n$ is therefore 
directly related to the fundamental physics  governed by the tunneling
action.   

Finally, the time-scale of the transition in terms of conformal time is
 \beq
  \widetilde{\beta}_n^{-1} = \eta_{s, n} - \eta_{i, n} = a(t_n)^{-1}\beta_n^{-1} 
  \, . 
 \eeq

\subsection{Quantum vacuum fluctuations}\label{vacuumwaves}

As a useful reference and comparison, we report here the spectrum
of the gravitational waves generated via fluctuations in vacuum
of the gravity field during inflation. 
The wave equation is the homogeneous version of
(\ref{gravitywaveseqmot}), with $\pi^T_{ij} = 0$.
Its solution leads to the spectrum for superhorizon modes
 \beq \label{vacuumspectrum}
  P^{\text{supQ}}_h= {2 \ov \pi^2} \left({H \ov M_{\text{Planck}}} \right)^2
 \eeq
 using the Bunch-Davies vacuum. $H$ is the Hubble parameter
 during inflation.



\end{document}